\def \date         {\ifcase\month \message{zero} \or
                    January \or February \or March \or April \or May \or June 
                    \or July \or 
                    August \or September \or October \or November \or 
                    December \fi
                    \space\number\day, \number\year}
\def \eg           {{e.g.}}
\def \etal         {{et~al. }}
\def \h2         {\hbox{H$_2$}}

\def \IRAS         {\hbox{{\it IRAS\ }}}
\def \kms          {\hbox{km$\,$s$^{-1}$}}
\def\approxlt{\lower.2em\hbox{$\buildrel < \over \sim$}}
\def\approxgt{\lower.2em\hbox{$\buildrel > \over \sim$}}

\def \ls           {\hbox{L$_{\odot}$}}
\def \Lsun         {\hbox{L$_{\odot}$}}           % Solar luminosity

\def \ms           {\hbox{M$_{\odot}$}}           % Solar mass
\def \Msun         {\hbox{M$_{\odot}$}}           % Solar mass

\def \ha           {H$\alpha$} 
\def \kms          {\hbox{km$\,$s$^{-1}$}}

\documentstyle[10pt,aas2pp4]{article}
\begin{document}

\title{CO in Stephan's Quintet: First Evidence of 
Molecular Gas in the Intragroup Starburst}

\author{Yu Gao\altaffilmark{1,2,3} and Cong Xu\altaffilmark{1}}

\altaffiltext{1}{Infrared Processing and Analysis Center, 
Jet Propulsion Laboratory, Caltech 100-22, Pasadena, CA 91125}
\altaffiltext{2}{Department of Astronomy, University 
of Toronto, 60 St. George Street, Toronto, ON M5S 3H8, CANADA}
\altaffiltext{3}{Laboratory for Astronomical Imaging, Department of Astronomy,
        University of Illinois, 1002 W. Green Street, Urbana, IL 61801}

\received{\date}
\accepted{ }

\begin{abstract}

We present the first interferometric evidence of CO(1-0) emission
in the intragroup starburst (IGS) region of HCG~92
(Stephan's Quintet). BIMA's large primary beam covers fully both 
the IGS region and the dominant group member galaxy Seyfert 
NGC~7319, as well as partially the intruder galaxy NGC~7318B.
CO emission of $\approxgt 4\sigma$ is detected
in all of them. The detection of the CO emission
associated with the IGS is further supported by 
comparisons with observations in other wavebands (e.g., HI, mid-IR, \ha).
Assuming the standard conversion factor, 3.6--8.0$\times 10^8~\Msun$ 
molecular gas is found to be associated with the IGS. 
Confining to the region of the CO emission peak,  
a relatively high star-formation efficiency of 8.6$\Lsun/\Msun$ is
derived for the IGS, which is nearly comparable to that of local starburst
galaxies (e.g., M82).

\end{abstract}


\twocolumn
\section{Introduction}

Stephan's Quintet (SQ), initially
discovered 120 years ago and numbered 92 in the catalog of 
the Hickson Compact Groups (HCG~92, Hickson 1982),
is one of the most famous and best studied 
galaxy groups (e.g., Moles, M\'arquez, \& Sulentic 1998). Recent discovery
of a prominent intragroup starburst (IGS) in the intragroup
medium (IGM) of SQ, far away from centers
of member galaxies, by Infrared Space Observatory (ISO) mid-IR and 
ground-based \ha~ and near-IR observations (Xu, Sulentic, 
\& Tuffs 1999), demonstrates that a starburst can be triggered by a high
speed ($\sim 1000$ \kms) collision between an intruder galaxy and cold
IGM. 

Despite the richness of the data from
multi-wavelength observations, little CO observations
have been obtained to probe the molecular gas content 
of SQ. At present, it is unknown whether 
there is any molecular gas readily detectable in the
IGS region. Yun \etal (1997) detected CO from the dominant member 
Seyfert NGC~7319, but missed the IGS due to 
the limited field-of-view.
Even the CO observations of NGC~7319 are controversial, 
off by a factor of 10 in the estimated molecular gas mass
(Leon, Combes, \& Menon 1998; Verdes-Montenegro \etal 1998, V98; 
Yun \etal 1997). Although SQ is one of
the HI deficient galaxy groups surveyed by single dish 
(Williams \& Rood 1987), several HI gas concentrations have been detected
in IGM of SQ, especially in the IGS region where 
$\sim 10^9 \ms$ HI is detected (Shostak, Sullivan, \& Allen 1984, S84; 
Williams, Yun, \& Verdes-Montenegro 1999, 2000 in preparation, W00). 
Should the molecular gas mass of the IGS be close to the HI mass, 
the detection of CO with current instrumentation would be 
straight forward even at the 
distance of 89 Mpc ($H_0=75~\kms$~Mpc$^{-1}$, $1'' \sim 0.4$~kpc).

Since stars are born from the cold giant molecular clouds (GMCs),
CO observations can provide strong constraints to the
physical mechanism of the IGS, which is still poorly understood
(Xu et al. 1999). It is also of great interest to see whether
molecular IGM can survive in the rather hostile
environment in SQ, where large scale shocks are 
happening (Pietsch et al. 1997). Previously, the only detections of
the IGM molecular gas are in the group of M81
(Brouillet et al. 1992; Walter \& Heithausen 1999), where 
little star formation is detected (Henkel et al. 1993). 
Motivated by these considerations,
we carried out the CO observations presented in this letter.

\section{Observations}

The CO(1-0) observations of SQ have been carried out with 
the Berkeley-Illinois-Maryland Association (BIMA) 10-element
millimeter array at Hat Creek, northern California. 
Two compact C array tracks (synthesized beam $\sim 6''$) under 
good weather condition (a typical system temperature 
$\approxlt 300$~K and stable phase) were conducted in Oct. 1998. 
Judging from H$\alpha$ and mid-IR observations (Xu \etal 1999), 
the source size of IGS may be only a few arcseconds 
although weak emission extends to $\sim 20''$. 
Ideally, long baseline B array imaging at 2$''$ 
is desired in order to study gas properties on a kpc scale. 
Three B array tracks were therefore 
observed in Feb./March 1999. However, only one of them was 
under good weather 
condition whereas the other two were marginal. Thus, 
more B array observations have been obtained in March 2000. 

We intended to detect CO emission
in the IGS region as well as to confirm 
the unusual CO features in NGC~7319 detected by Yun \etal (1997).
We have thus pointed the BIMA's primary beam (FWHM$\sim 110''$)
to cover NGC~7319, IGS and part of the intruder galaxy NGC~7318B. 
We have tuned the SIS receivers to 112.9356 GHz, 
corresponding to 6200~\kms, and configured the correlator to cover a velocity
range of 1600~\kms~ at a spectral resolution of 
$\sim 8$~\kms~(3.1 MHz). 
This is because there are at least two HI kinematic systems 
in IGS region (S84; W00) and 
NGC~7318B is at a velocity 1000~\kms~ lower. We have smoothed the datacube
to 40 \kms ($\sim 15$~MHz) so that
the sensitivity achieved allows us to identify 
any features similar to GMCs complex 
associations observed in ISM of nearby galaxies.

2203+317 was observed as a phase calibrator and
Uranus/Mars for absolute flux
calibrations. We have also applied a Gaussian taper 
to down weight long baseline visibilities to compromise the 
atmospheric decorrelation. The cleaned maps 
from all observations combined have
a final synthesized beam of 8.8$''\times 8.3''$.

\section{Results}

In Fig.~1, we compare the integrated CO intensity contours 
with the ISO mid-IR, R-band CCD and the continuum subtracted 
high velocity (6600 \kms) \ha~ images of SQ. Remarkably, 
we have detected the CO emission from the IGS 
with the strongest $\sim 4\sigma$ detection right at the 
mid-IR peak (Fig.~1a). Also shown are the
extracted CO spectra from the datacube for all likely
detected features (Fig.~1b). On the other hand, 
the strongest \ha~ and optical features correspond to
a weaker marginal CO feature (nearly 3$\sigma$), rather than 
the CO peak (Fig.~1c and Fig.~1d).
The offset between the CO peak and the \ha~ and optical peaks
could be due to the dust extinction usually associated with
CO emission (e.g., in the overlap region of Arp~244, Gao et al. 2000).
Examination of the CO spectra
and velocity channel maps indicates that the weak CO emission might 
have been associated with 3 or 4 velocity components: high 
velocity between
6500 and 6750 \kms, medium velocity CO around 6200 and at
6020 \kms, and possibly the weakest low velocity CO at 5750 \kms.
All these components (except 6200~\kms) are at 
exactly the same velocity
ranges of the HI complexes revealed from the VLA observations
although the 5700~\kms~ component seems to be in south of IGS region
(S84; W00). There might be \ha~ emission
in IGS at 5700~\kms~ although this can also be due to the
contribution of \ha~ emission from the 6000~\kms~ component (Xu \etal 1999).
The 6200~\kms~ CO emission appears to be a new component. Yet the reality 
needs additional observations to check as some 
negatives also appeared at this velocity in NGC~7319. 
Of course, high sensitivity data are obviously required to better
study the molecular gas distribution and kinematics in this very interesting
IGS region. 

We confirm the result of Yun \etal (1997)
that the CO distribution in the Seyfert galaxy NGC~7319 is unusually 
concentrated in two CO complexes. The dominant CO in NGC~7319 resides in the
dusty regions in northern tidal features (Fig.~1c)
which show distinct arclike \ha~ emission (Fig.~1d).
The nuclear CO in 
NGC~7319 is elongated almost perpendicular to the inclined 
stellar disk and similar morphology in \ha~ may have also
been observed. We find that the nuclear CO peaks ($\sim 6 \sigma$)
almost at the same position ($<1''$) as the newly obtained VLA 20cm 
radio continuum peak (Condon \etal 2000, in preparation),
which could be the Seyfert nucleus. But Yun \etal (1997) claimed 
this nuclear CO feature to be 2~kpc ($\sim 5''$) from the 
nucleus. Obviously, the stellar disk of NGC~7319 is
deficient of both molecular gas and on-going star formation
except for the inner several kpc nuclear region. Furthermore,
the CO emission is almost exclusively at the highest velocity
of the entire system, between 6400 and 6800~\kms~ in the nuclear
region and narrowly peaks between 6750 and 6870~\kms~ in the
northern dominant CO complex (Fig.~1b).  

Other two CO features appear also to be $\sim 4 \sigma$ level
in the integrated CO intensity map. One is near the center of 
the BIMA's primary beam (shown as the black circles in Fig.~1)
and is probably not real judging from the channel maps and the
CO spectrum (labeled as `Beam Center') in Fig.~1b. This is 
because no signal at $\approxgt 3\sigma$ level is apparent in any channel 
at this location except at 5700 \kms.
The other, just outside the BIMA's beam, is mostly 
at a velocity of 6250 \kms~ and possibly at 5700 \kms. This is 
likely to be associated with some CO clouds near nuclear region of
NGC~7318B as both the location and velocities might
have matched. Nevertheless, 
CO emission is distributed offset from the nucleus, though very 
close, and most of CO, if real, appears to be at
high velocity of 6250~\kms~ rather than 5700~\kms. 

The total CO flux detected from the Seyfert is
$38.7 \pm 2.0$~Jy~\kms~ (with only 1/4 of this
from the nuclear region), thus a molecular gas mass of
M(H$_2)=3.6\times10^9\ms$ using the
standard CO-to-H$_2$ conversion:
$M(H_2) = 1.177\times10^4 S_{\rm CO} dV d_{\rm L}^2$~\ms, 
where $S_{\rm CO} dV$ is the CO flux and $d_{\rm L}$ is 
the distance in Mpc. We have observed
much more extended CO emission in both CO features
as compared to what have been detected by Yun \etal (1997). 
In particular, the dominant CO has a CO extent of more than 10 kpc.

The total CO emission from the IGS region is
more uncertain given the weak features detected. Our estimated
CO flux for clumps immediately surrounding the $4\sigma$ peak is 
$\sim$4.0~Jy~\kms, but the weak CO extension towards north appears to
have several more clumps with similar or more CO flux of $\sim$5.0~Jy~\kms.
If all these are real, the total CO flux can be as 
large as 9.0$\pm1.6$~Jy~\kms. Therefore, the molecular gas mass
closely associated with this IGS region is
$\sim 3.6\times10^8\ms$, but can be as large as 8.0$\times10^8\ms$.
This is nearly comparable to the $10^9\ms$ of HI
detected in this IGS region (S84; W00). 

Previous observations
did not detect any CO emission from the prominent interacting pair
NGC~7318A/B (Yun \etal 1997; V98). 
Our $\sim 4\sigma$ detection gives a total CO flux
$\sim 3.1$~Jy~\kms, or a molecular gas mass of 
2.8$\times10^8\ms$ for NGC~7318B even though the feature is just beyond
the BIMA's beam. In comparison, V98 
listed an upper limit of 12.6$\times10^8\ms$.
Overall, the spatial coincidence of weak CO features with the optical,
\ha, mid-IR and HI emission as well as the matched velocities suggests 
that most of CO emission could be real.

\section{Discussion}

\subsection{Intragroup Starburst: Star Formation Efficiency and Triggering Mechanism}

The reliability of our CO detection of the IGS 
is strongly supported by two pieces of evidence:
\begin{description}
\item{(1)} The close correlation with the HI gas. Not only the 
{\it positions} of the CO signals coincide well with the peaks of
the HI components (S84; W00), but also the
velocity components of the CO emission agrees well with those of the HI gas
(Fig.~1b).
\item{(2)} The CO emission peaks right at the place where the mid-IR
emission of the IGS peaks (Fig.~1a), which is indeed expected
given the close relations of both to massive star formation.
\end{description}

The star formation efficiency (SFE), defined as the
IR luminosity to the molecular gas mass ratio, is a good
indicator of the starburst strength.
The mean values of SFE for normal galaxies is about 4~$\Lsun/\Msun$,
$11~ \Lsun/\Msun$ for nearby starbursts and only in 
luminous (and ultraluminous) infrared galaxies (LIGs), 
SFE $\approxgt 20 \Lsun/\Msun$ (Sanders \& Mirabel 1996), similar to what has
been found in the star forming cores of GMCs. 

For IGS of SQ, there is no IR luminosity available.  
Therefore we carried out the following estimates: 
(1) Using the mid-IR map (Xu et al. 1999)
to scale the IR luminosity of the whole group. The total
IR luminosity L$_{\rm IR}$ for the entire group, based on
\IRAS measurements, is $\approxlt 3.61\times10^{10}\ls$ 
since both 12 and 25~$\mu$m are upper limits.
From Table 1 in Xu \etal (1999), the mid-IR emission of the IGS 
is about 9\% of the total mid-IR emission in the entire field. 
Assuming that the total IR emission of IGS is 
9\% of the total IR emission of SQ, we derive an IR luminosity of
$3.2\times 10^9 \ls$ for IGS. 
(2) Using radio continuum emission at 20cm to derive 
the far-IR emission based on the well-known correlation
between the two since 
$q=2.35=log[F_{\rm FIR}/(3.75\times10^{12} Hz)/f_{\nu}(1.49GHz)]$
(\eg, Helou, Soifer, \& Rowan-Robinson 1985; Xu \etal 1994).
We estimated the radio continuum emission of the IGS region 
(Condon \etal in preparation; also van der Hulst \& Rots 1981)
to be $f_{\nu}(1.49GHz) \sim 1.4$~mJy. We thus obtained
the IR luminosity of $3.1\times10^9 \ls$ for the IGS, 
same as estimated from the mid-IR map.

Hence, the SFE=$8.6 \Lsun/\Msun$ for IGS 
(if the far-IR emission of this region is mostly from
the 4$\sigma$ peak CO emission) or $\sim 4 \Lsun/\Msun$ (if all weak 
CO features are real and far-IR emission is spread over the entire
IGS region). Given the localized nature of
starburst and our weak CO detection, we tend to adopt 
$8.6 \Lsun/\Msun$ as the SFE of the IGS, which
is indeed comparable to that of local starburst galaxies. 

It is tempting to speculate on the cause of such a high SFE 
in a rather abnormal environment of the IGM. Generally, 
high SFE is associated with high molecular
gas concentrations in advanced LIG mergers (\eg, Scoville
\etal 1991; Gao \& Solomon 1999). For most 
starburst galaxies (including LIGs), the high gas density is
due to interaction induced gas infall
into the inner few 100 pc of the nucleus during
the merging (Sanders and Mirabel 1996).
This is certainly not the case for the IGS in SQ,
which is far-away from any galaxy centers. On the other hand,
as noticed by Xu et al. (1999), the IGS is located in
a hole of the X-ray emission, and comparison between our CO map and 
the new HI map of W00 with the X-ray map of
Pietsch et al. (1997) suggests that the highly concentrated 
cold gas at the place of the IGS could be currently undergoing
through a squeezing (compression) by the surrounding hot
gas. This may increase the cold gas density significantly and 
trigger the high SFE starburst in IGM. 

The scenario suggested above is different from the 
one published by Jog \& Solomon (1992), in which
a starburst occurs when the preexisting GMCs in  
the colliding disks of LIGs are compressed by shock-heated
hot gas (caused by HI clouds collision) that immediately surrounds them.
Because no X-ray emission is detected in the IGS region,
such shock-heated hot gas apparently does not exist {\it within} the
region (or already cooled). Therefore,
instead of the individual GMCs being directly squeezed by hot gas,
it is more likely that the whole region is being squeezed
to a higher density. The starburst is then caused by the high
density of the cold gas, rather than being a direct consequence of the
compression of GMCs as in the Jog \& Solomon (1992) scenario.

\subsection{Molecular Gas in SQ: Status Report}

We have detected larger CO flux in NGC~7319 than the reported
single-dish measurements. While the NRAO 12m's FWHM beam 
of 55$''$ should cover
almost all CO features in NGC~7319, the IRAM 30m's $22''$ beam
would miss almost entirely the dominant CO in the 
north (assuming pointings were good and centered on the
Seyfert nucleus). Yet, the
30m (Leon \etal 1998) obtained larger CO flux of 
$\sim 33$~Jy~\kms~ whereas the 12m (V98)
gave a CO flux of $\sim 18$~Jy~\kms~ for NGC~7319 (compared to 
ours of 38.7~Jy~\kms~). 
Checking the published single-dish CO spectra, both seem to be 
marginal and large uncertainties in the CO flux
are expected. While the large molecular gas mass of
9.3$\times 10^9$\ms~ estimated by Leon \etal (1998) might be
an overestimate (they introduced a large scaling factor
to account for the large CO source size missed by the 30m beam),
the molecular gas mass of 1.5$\times 10^9$\ms~ listed 
in V98 ($8.6\times10^8\ms$ in Yun \etal 1997)
seems to be too low to compare with the total HI gas mass in SQ.
We seem to conclude that a more reasonable molecular gas mass 
of NGC~7319 is $\approxgt 3.6\times 10^9$\ms.

Although the molecular gas mass of IGS is more uncertain, 
a few times of 10$^8\ms$
is more massive than the GMC associations in nearby galaxies.
This is also much more massive than the molecular complexes
discovered in the M81 group (Brouillet et al. 1992; 
Walter \& Heithausen 1999), which has a mass of 10$^6$--10$^7\ms$.
Most of the cold gas (atomic and molecular) concentrated
in the IGS region seems to be in the 6600 \kms~ component,
associated with the preshocked IGM (Allen and Sullivan
1980; S84). This gas is perhaps stripped
from NGC~7319 a few 10$^8$ years ago (S84;
Moles et al. 1997). Because it is usually difficult to strip the
molecular gas (deeper inside the inner disk than the HI gas),
it is more likely that the molecular gas we detected in the
IGS region is condensed locally from the HI gas as the
CO peaks right at the HI peak,
presumably in connection with the squeezing process suggested
above. We cannot rule out the possibility that the molecular gas 
in IGS region is stripped from NGC~7319, however, given the 
large distance of the largest CO complex away from the galaxy center
(Fig.~1a). In that case, the
atomic--molecular gas relation might be 
completely opposite, namely that some HI gas
in IGS might be a result of the photo-dissociation (due to the IGS)
of the stripped molecular gas.

Since there are no consistent single-dish observations, 
we do not know how much of the
extended CO emission could still be missed due to the
zero-spacing problems inherent to an interferometer.
Therefore, the sum of the 
molecular gas mass from all likely detected CO features should be 
a lower limit to the true total molecular gas content of the entire group.
We thus obtain a total molecular gas mass of $\approxgt 4.7\times10^9\ms$ 
for SQ (mostly in NGC~7319), which is about
half of the total atomic gas mass in SQ (Williams \& Rood 1987)
exclusively distributed {\it outside} member galaxies in IGM  
(S84; W00). 
Our CO detection in IGS region of SQ appears to indicate that the molecular 
gas could also be an important component in the {\it cold} gas of the IGM,
$\approxgt 30\%$ of the HI gas mass, which can directly 
provide the fuel for the IGM starburst.

\acknowledgments
Drs. L. Verdes-Montenegro and B.A. Williams are thanked for providing
the VLA HI maps of SQ before publication. 
This work was supported by NASA grant for ISO data analysis.
This research has made use of the NASA/IPAC Extragalactic Database (NED) 
which is operated by the Jet Propulsion Laboratory, California Institute of 
Technology, under contract with the National Aeronautics and
Space Administration. 
Y.G. and C.X. are supported by the Jet Propulsion Laboratory,
California Institute of Technology, under contract with NASA.

\clearpage
 
\onecolumn
\begin{figure}
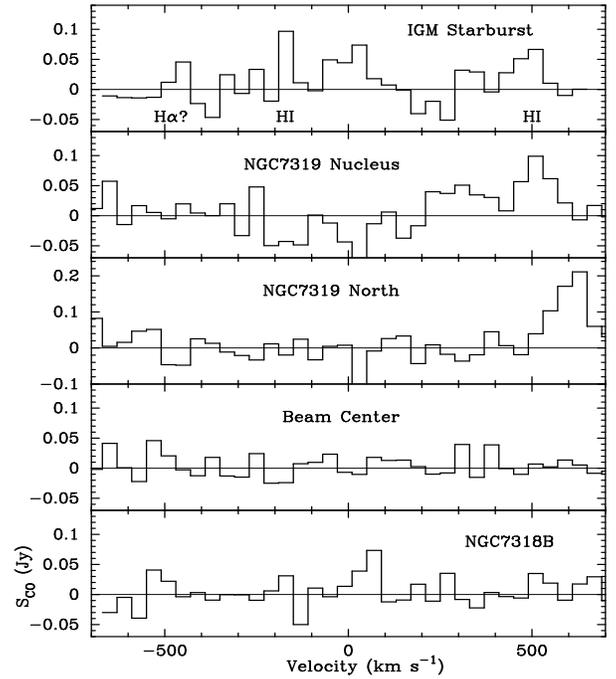


\figcaption{\small (a) (top left) ISO mid-IR image overlaid with 
BIMA CO(1-0) integrated intensity contours. Contours start at 
2.4, 3$\sigma$ and increase by 1$\sigma=1.1$~Jy/Beam~\kms. 
The dark circles indicate the primary beam of 110$''$ (FWHM). 
(b) (top right) Various CO spectra extracted from the
datacube for those possibly detected 
CO features. Zero velocity refers to
6200~\kms. HI emission peaks and possible \ha~ components are marked
in the panel for `IGM starburst'.
(c) (bottom left) Comparison of CO contours (same as Fig.~1a) with 
R-band CCD image. (d) (bottom right) Same CO contours
are compared with the high velocity 6600 \kms~ \ha~ image.}
 
\plotfiddle{fig1a.ps}{3.5in}{0}{50}{50}{-270}{-80}
\plotfiddle{fig1b.ps}{3.5in}{0}{46}{42}{15}{235}
\plotfiddle{fig1c.ps}{3.5in}{0}{50}{50}{-270}{210}
\plotfiddle{fig1d.ps}{3.5in}{0}{50}{50}{-10}{480}

\end{figure}

\end{document}